\documentclass[onecollarge]{svjour2}       
\smartqed  
\usepackage{graphicx}
\usepackage{color}
\usepackage{mathptmx}      
\usepackage{amsfonts}
\newcommand{\rd}[1]{#1}

\newcommand{\calB}{\mathcal{B}}
\newcommand{\trho}{\tilde{\rho}}

\journalname{}
\begin{document}

\title{Jamming transition in kinetically constrained models with \rd{reflection} symmetry 
}


\author{Hiroki Ohta \and Shin-ichi Sasa 
}


\institute{Hiroki Ohta \at
              Yukawa Institute for Theoretical Physics, Kyoto University, Kyoto 606-8502, Japan \\             
              \email{Hiroki.Ohta@nbi.dk}             \\
            \emph{Niels Bohr Institute/NBIA, University of Copenhagen, Copenhagen 2100, Denmark}  
           \and
           Shin-ichi Sasa \at
             Department of Physics, Kyoto University, Kyoto 606-8502, Japan
}

\date{}

\maketitle

\begin{abstract}
A class of kinetically constrained models with \rd{reflection} symmetry 
is proposed as an extension of the Fredrickson-Andersen model.
It is proved that the proposed model on the square lattice 
exhibits a freezing transition at a non-trivial density. 
It is conjectured by numerical experiments that the known mechanism of the singular behaviors 
near the freezing transition in a previously studied model (spiral model) 
is not responsible for that in the proposed model.

\end{abstract}

\section{Introduction}


Athermal particles with repulsive interactions exhibit 
rigidity with an amorphous structure 
when the density of the particles is higher 
than a critical value. Such a phenomenon is referred to as 
a {\it jamming transition}, and elucidation of the nature of jamming transitions 
is a challenging problem in statistical physics 
\cite{Liu,Saaloos,Olsson,Hatano,Otsuki,Dauchot1,Dauchot2,Dauchot3}. 
Thus far, the replica theories and the related cavity methods 
for equilibrium glass models have provided insights into
jamming transitions as well as glass transitions 
\cite{Zamponi,Mezard}. As a different approach, 
kinetically constrained  models (KCMs), 
which were investigated with a physical picture 
that glassy dynamics is purely kinematic \cite{Fredrickson,Kob,Sollich},  
have been considered for understanding jamming transitions \cite{TBF}. 
A particular property of KCMs is the absence of singularities 
in the equilibrium properties. 
Nevertheless, dynamical phase transitions in KCMs on Bethe lattices 
have been known to be strongly connected to 
a certain dynamical phase transition called a freezing transition 
in equilibrium glass models \cite{Semerjian}.
Also, recent studies have attempted to reveal the relationship between 
KCMs and glass-forming materials \cite{Dauchot,Garrahan}.


Let us review theoretical studies on KCMs in brief.
Here, we generally call dynamical phase transitions in KCMs 
freezing transitions, which mean the transition 
from an equilibrium phase to a frozen phase where an infinite number of 
particles are at rest as a result of blocking by other particles.
The first proof for the existence of a 
freezing transition was presented for the Fredrickson-Andersen (FA) model 
and then also the Kob-Andersen model on a Bethe lattice \cite{SBT,TBF0}. 
However, for these models in finite dimensions, 
it has been shown that there are no true phase transitions \cite{TBF0,Toninelli1}, 
even though the numerical simulations have shown an indication of a transition.
Then, it has been proved that a KCM, which is called spiral model, 
exhibits a freezing transition in two dimensions \cite{Toninelli2,Toninelli3}, 
leading to the concept of universality classes (we call that of the spiral model {\it spiral} class) 
for finite-dimensional freezing transitions.

However, the meaning of such a concept of universality classes
for KCMs, in particular, in finite dimensions has not been deeply understood
because the spiral model is somewhat an ideally simple case that does not have \rd{reflection} symmetry, 
and the classes different from the {\it spiral} class have not been known yet. 
Thus, the previous studies have attempted to estimate the universality classes 
in the other models: a finite-size scaling analysis using a special boundary condition 
for the knight model is consistent to the {\it spiral} class \cite{Toninelli0}, 
and the robustness of the universality against two (periodic and filled) 
boundary conditions for a wider class of models 
showing the freezing transition in finite dimensions has been investigated, which leads 
that the universality class of investigated models in two dimensions also 
looks consistent to the {\it spiral} model \cite{Schwarz} 
although the numerical data are not as conclusive as that with using a special boundary condition \cite{Toninelli0}.
Therefore, it is a significant challenge to present a KCM that 
exhibits a freezing transition but does not belong to the {\it spiral} class. 
We attempt to solve this problem.


In particular, we focus on KCMs with reflection symmetry,  
which may be conjectured as qualitatively different 
from KCMs in the {\it spiral} class owing to the symmetry of the models. 
We also search for models, which in their mean-field cases such as the models on 
the infinite dimensional lattice and Bethe lattices are straightforwardly well-defined. 
Mean-field analysis for such cases possibly acts as a starting 
point for a theory on determining the universality, 
as we have learned from the renormalization group analysis for equilibrium critical phenomena.
With this background, the strategy in this paper is to first remind us of the FA model, which is 
straightforwardly well-defined on Bethe lattices but does not show 
any freezing transitions in finite dimensions.
Then, by adding the other constraints to the FA model, we propose a KCM with \rd{reflection} symmetry, 
for which the existence of a freezing transition can be proved on the square lattice. 
Further, we conjecture by numerical experiments that the proposed model 
does not belong to the {\it spiral} class.
 

\section{Model} \label{model}

\begin{figure}
\centering
\includegraphics[width=3.0cm]{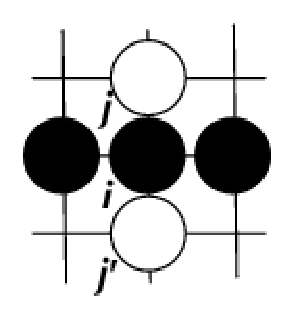}
\includegraphics[width=7.0cm]{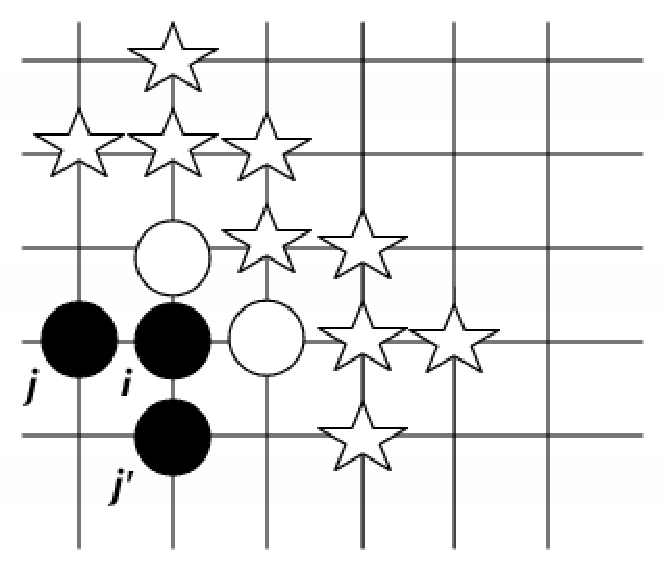}
\caption{
Two configurations satisfying $\sum_{j\in {\rm B}_i} \sigma_j=2$. 
In the configuration to the left,
the particle at site $i$ is constrained, because 
$f_j$ and $f_{j'}$ are at most $1$. 
In the configuration to the right, the particle at site $i$ is 
not constrained only when there are no particles at the sites marked 
with the star symbol. The particle at site $i$ becomes 
constrained when a particle is placed at the sites marked 
with the star symbol.}
\label{block}
\end{figure}


Let $\Lambda$ be a square lattice consisting of sites 
$i \equiv (i_x,i_y)$, where $i_x$ and $i_y$
are integers satisfying $1 \le i_x \le L$ and $1 \le i_y \le M$, 
where $M=L$ is assumed unless otherwise specified. 
We define the occupation variable $\sigma_i \in \{0,1\}$ 
at each site $i \in \Lambda$, 
and assume a Hamiltonian 
\begin{eqnarray}
H(\sigma)= \sum_{i \in \Lambda }\sigma_i,
\label{hamil}
\end{eqnarray}
where we express $\sigma \equiv (\sigma_i)_{i \in \Lambda}$
collectively. Concretely, $\sigma_i=1$ represents a state that 
a particle occupies site $i$, while there is no particle 
at site $i$ when $\sigma_i=0$. In the argument below, 
we assume that particles are filled outside of the system\rd{, 
which we call the filled boundary condition,} unless otherwise specified.
\rd{In fact, in numerical experiments at section 4.1, 
we use the periodic boundary condition where sites $\{L,y\}$ and $\{x,L\}$ 
are connected to $\{1,y\}$ and $\{x,1\}$ for any $x$ and $y$, respectively; 
also, at section 4.2, we use the half-filled boundary condition that 
particles are filled in the left and right region of the outside of the system, 
while no particles exist in the top and bottom region of the outside of the system.}
We consider a Markov stochastic process with a transition ratio 
$R(\sigma \to \sigma')$ for $\sigma \not = \sigma'$.
Let $P(\sigma, t)$ be the probability distribution of a particle configuration $\sigma$ at time $t$. 
The time evolution of $P(\sigma,t)$ obeys
\begin{eqnarray}
\partial_t P(\sigma,t)=\sum_{\sigma'\not = \sigma}
[R(\sigma' \to \sigma)P(\sigma',t)
-R(\sigma \to \sigma')P(\sigma,t)].
\label{mas}
\end{eqnarray} 
In this work, we study a constrained Glauber dynamics, 
where the transition ratio $R(\sigma \to \sigma')$ is written as
\begin{equation}
R(\sigma\to \sigma')
= \sum_{i} 
\delta(\sigma', F_i \sigma)
r(\sigma\to \sigma'){\mathcal C}_i(\sigma).
\label{trate}
\end{equation}
We explain the right-hand side of the equation in order.  
First, $F_i$ is the creation and annihilation operator described by
\begin{equation}
(F_i\sigma)_j=(1-\sigma_i)\delta_{ij}+\sigma_j(1-\delta_{ij}).
\end{equation}
The term $\delta(\sigma', F_i \sigma)$ represents
a rule that the transition is caused by a particle creation or annihilation
at each site. The term $r(\sigma\to \sigma')$ is given as
\begin{eqnarray}
r(\sigma\to \sigma')
=\min
\left[ 1,\exp \left( \frac{H(\sigma)-H(\sigma')}{T} 
\right) \right],
\end{eqnarray}
where it satisfies 
\begin{eqnarray}
\frac{r(\sigma\to\sigma')}{r(\sigma'\to \sigma)}
=\exp\left(-\frac{H(\sigma')-H(\sigma)}{T}\right),
\end{eqnarray} and the Boltzmann constant is set to unity. 
Finally, with regard to the function ${\mathcal C}_i(\sigma)$,
${\mathcal C}_i(\sigma)=0$ specifies a set of configurations 
for which particle creation and annihilation at site $i$ is prohibited, 
and ${\mathcal C}_i(\sigma)=1$, for the other configurations. 
We also assume that ${\mathcal C}_i(\sigma)$ is independent of $\sigma_i$.


It should be noted that the canonical distribution with 
the Hamiltonian (\ref{hamil}) is the stationary solution 
of (\ref{mas}) with (\ref{trate}) because the 
transition ratio (\ref{trate}) satisfies the detailed 
balance condition. Therefore, there are no equilibrium phase 
transitions in the system. Nevertheless, it has been shown that
there is a certain dynamical transition (freezing transition) 
for an appropriately selected ${\mathcal C}_i(\sigma)$. 
In all examples known thus far, 
${\mathcal C}_i(\sigma)$ are described by using the oriented structure of the lattice
\cite{TBF,Toninelli2,Toninelli3,Toninelli0,Schwarz}. 
In this paper, we generally assume that ${\mathcal C}_i(\sigma)$ 
is a \rd{reflection}-symmetric function satisfying ${\mathcal C}_i(\sigma)$ = ${\mathcal C}_i({\mathcal P}_i^w\sigma)$ 
for $^\forall i$ and any possible $w$ where \rd{a configuration $\mathcal{P}_i^w\sigma$ is 
given by the reflection of $\sigma$ with respect to the axis with angle $w$ from the $x$ axis.}
Note that $w$ takes $0$, $\pi/4$, $\pi/2$, and $3\pi/4$ in the case of the square lattice\rd{: 
for example, $\mathcal{P}_i^0\sigma_{i_x,i_y+d}=\sigma_{i_x,i_y-d}$ 
and $\mathcal{P}_i^{\pi/4}\sigma_{i_x-d,i_y+d}=\sigma_{i_x+d,i_y-d}$ for any integer $d$, 
and also in order to keep consistency in the cases of the filled and half-filled boundary condition, 
there are assumed to be sites on the infinite square lattice 
in the outside of the system.}

From now, we explain our selection of ${\mathcal C}_i(\sigma)$. We first define 
\begin{eqnarray}
f_i \equiv \sum_{j \in {\rm B}_i} \delta(\sigma_j,0)
\left[ \prod_{\ell \in {\rm B}_j} \delta(\sigma_{\ell},0) \right],
\end{eqnarray} where \rd{${\rm B}_i$ is a set of the nearest neighbor sites of site $i$ and} 
$\delta( m,n )$ represents Kronecker's delta function. 
$f_i$ represents the number of empty sites 
$j$ in ${\rm B}_i$ such that there are no particles next to any site $j$. 
From this definition,  we find $f_i=0$ for site $i$ with a particle.
Then, we set ${\mathcal C}_i(\sigma)=0$ 
when $\sum_{j \in {\rm B}_i} \sigma_j \ge k $ and
\begin{equation}
\sum_{j \in {\rm B}_i} \Theta(f_j-c/2) < c/2,
\label{ocon}
\end{equation} where $c$ is the connectivity of the lattice, $0\le k \le c$, and $\Theta(x)$ is a 
step function such that $\Theta(x)=1$ for $x\ge 0$, otherwise $\Theta(x)=0$. 
\rd{It should be noted that $c=4$ for the square lattice. 
For the square lattice, one can immediately find that } 
there are no freezing transitions for all the values of $k$ 
without condition (\ref{ocon}) because these are identical to the FA models \cite{TBF0,Toninelli1}. 
Moreover, the cases without condition (\ref{ocon}) 
are identical to those with condition (\ref{ocon}) for $k=3,4$. 
On the other hand, owing to condition (\ref{ocon}), 
the behaviors for the cases of $k=0,1,2$ become nontrivial. 
We explain condition (\ref{ocon}) with $k=2$ through the following examples. 
In this paper, we consider only this simple and nontrivial case with $k=2$, 
and we say that {\it the particle at site $i$ is constrained} when ${\mathcal C}_i(\sigma)=0$ and $\sigma_i=1$.
\begin{figure}
\centering
\includegraphics[width=8.0cm]{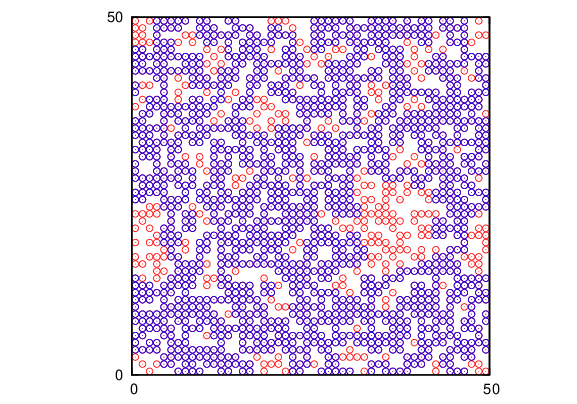}
\caption{(Color online) A particle configuration at an initial condition 
with $L=50$, the periodic boundary, and $\rho=0.69$: 
Blue points are frozen particles and red points are unfrozen particles.}
\label{image}
\end{figure}

First, one can find easily that a particle at site $i$ is constrained 
if $\sum_{j \in {\rm B}_i} \sigma_j = 3$ and $4$. 
More complicated situations occur at $\sum_{j\in {\rm B}_i} \sigma_j=2$. 
In figure \ref{block}, for the particle at site $i$, there are two cases
satisfying the condition $\sum_{j\in {\rm B}_i} \sigma_j=2$.
In the configuration to the left in figure \ref{block}, 
$f_j$ and $f_{j'}$ for the empty sites $j$ and $j'$ are at most $1$, 
leading to $\sum_{j \in {\rm B}_i} \Theta(f_j- 2) =0$. 
Therefore, in this case, the particle at site $i$ is constrained .
On the other hand, in the configuration to the right in figure \ref{block},
only when there are no particles at the sites marked with the star symbol,
$f_j$ and $f_{j'}$ for the empty sites $j$ and $j'$ are $2$, 
leading to $\sum_{j \in {\rm B}_i} \Theta(f_j- 2) =2$. 
In this case, the particle at site $i$ is not constrained.
When at least one particle is placed at the sites marked with the 
star symbol, $\sum_{j \in {\rm B}_i} \Theta(f_j- 2) =1$, 
which means that the particle at site $i$ is constrained as the result. 
Using similar considerations, one can also find that the function 
${\mathcal C}_i(\sigma)$ is independent of $\sigma_i$ 
and the \rd{reflection} symmetry condition 
${\mathcal C}_i(\sigma)={\mathcal C}_i(\mathcal{P}_i^w\sigma)$ holds
for $^\forall i$ and any possible $w$.




In this work, 
we focus on the relaxation behaviours of the system from an initial 
state, where sites are randomly occupied with a probability $\rho$. 
Moreover, as the simplest case, we consider the zero temperature 
limit $T \to 0$. The equilibrium state in this case involves 
no particles. Now, let us suppose that a particle configuration at $t=0$ contains 
a set of particles  constrained by only constrained particles and boundary sites. 
Such particles are referred to as {\it frozen particles}. 
As a reference, we show a configuration of frozen particles in figure \ref{image}.
Since frozen particles cannot annihilate, they can never reach the equilibrium state. 
On the other hand, when no frozen particles exist at $t=0$,
all the particles annihilate, and hence, the final state becomes
the equilibrium state. In this manner, whether a {\it freezing} transition, 
which is a phase transition between equilibrium states and 
such frozen states, occurs is determined by investigating the existence of frozen 
particles in initial configurations. 


\section{Existence of a freezing transition}\label{exis}
In this section, we prove that a freezing transition occurs at
a certain value $0 < \rho_{\rm c} < 1$ in the proposed model.
In concrete terms, we show that there is a density $\rho_{\rm l} > 0$, 
below which the probability of finding frozen particles at the bulk
in initial configurations goes to zero in the thermodynamic limit, 
while there is a density $\rho_{\rm u} < 1$, above which frozen particles can be 
found in initial configurations with probability one 
in the thermodynamic limit. By estimating such a lower bound $\rho_{\rm l}$ and 
an upper bound $\rho_{\rm u}$ of the transition point $\rho_{\rm c}$, 
we may conclude that $0 < \rho_{\rm l} \le \rho_{\rm c} \le \rho_{\rm u} <1 $. 

\subsection{Equilibrium states at $\rho<\rho_{\rm l}$}
Let us estimate a 
lower bound below which the probability of finding frozen particles at the bulk 
goes to zero in the thermodynamic limit. 
Specifically, let us consider whether the particle at centered site $(L/2,L/2)$, 
which is a representation of the bulk, is frozen. 
First, we assume that the clusters of frozen particles 
in the bulk are always constrained partly by the boundary sites
(See Appendix for the sketch of the proof).
Therefore, the minimum number of frozen particles 
including this centered site is more than $L/3$, 
because each frozen particle in one frozen cluster 
has to be located at a site within distance $3$ from 
another frozen particle by the definition, 
and also such a sequence of the frozen particles has to percolate from one boundary 
and another boundary with length $L$.

Concretely, let $(i^{(k)})_{k=1}^N$ be a sequence of $N$-sites such that
(0) $^\exists k$ satisfying $k=(L/2,L/2)$ 
(i) $\sigma_{i^{(k)}}=1$ for $^\forall k$, 
(ii) $ 0< |i^{(k)}-i^{(k+1)}| \le 3$ for $^\forall k$
(iii) $|i^{(k)}-i^{(k')}| >  3$ for $^\forall k$ and $^\forall k'\not = k\pm 1$ 
where the distance between site $i$ and $j$ is denoted by $|i-j|\equiv |i_x-j_x|+|i_y-j_y|$. 
Note that (ii) and (iii) lead to 
(iv) $i^{(k_1)}\neq i^{(k_2)}$ for $^\forall k_1$ and $^\forall k_2 \neq k_1$. 
For a given configuration
$\sigma$, we define a set of all such sequences, 
which is denoted by ${\cal D}_N(\sigma)$. From the fact mentioned above,
we can always find a sequence of sites in ${\cal D}_{L/3}(\sigma)$
if frozen particles including the centered site 
are present in the initial configuration $\sigma$, 
where $L/6$ is assumed to be an integer without loss of generality.
Thus, the probability $Q$ of finding a frozen particle at 
the centered site in initial configurations satisfies 
\begin{eqnarray}
Q\le{\rm Prob}({\cal D}_{L/3} \not = \phi).
\end{eqnarray}
Here, we also obtain  
\begin{eqnarray}
{\rm Prob}({\cal D}_{L/3} \not = \phi)\le{\rm Prob}(^\exists(i^{(k)}_*)_{k=1}^{L/3} ),
\end{eqnarray}
where $(i^{(k)}_*)_{k=1}^{L/3}$ satisfies conditions 
(0), (i), (ii), and (iv) that is weaker than (iii).
Further, by recalling the conditions (0) and (iv), we easily obtain 
${\rm Prob}(^\exists(i^{(k)}_*)_{k=1}^{L/3} )\le W^{L/6-1}$, 
where $W\equiv\sum_{i;0<|i-j|\le 3} {\rm Prob}(\sigma_{i}=1|\sigma_{j} = 1)
=24\rho$. These estimations give 
\begin{eqnarray}
Q \le (24\rho )^{L/6-1}.
\end{eqnarray}
Therefore, when $\rho$ is less than $1/24$, $Q \to 0$
in the thermodynamic limit. The obtained results can be applied 
to the sites which are sufficiently far from the boundaries.
We thus find a lower bound as $\rho_{\rm l}=1/24$.

\subsection{Frozen states at $\rho > \rho_{\rm u}$}
\begin{figure}
\centering
\includegraphics[width=6cm]{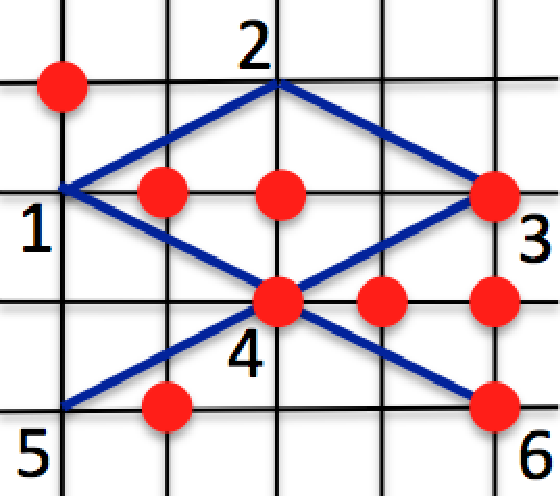}{\ }{\ }{\ }{\ }{\ }{\ }{\ }
\includegraphics[width=6cm]{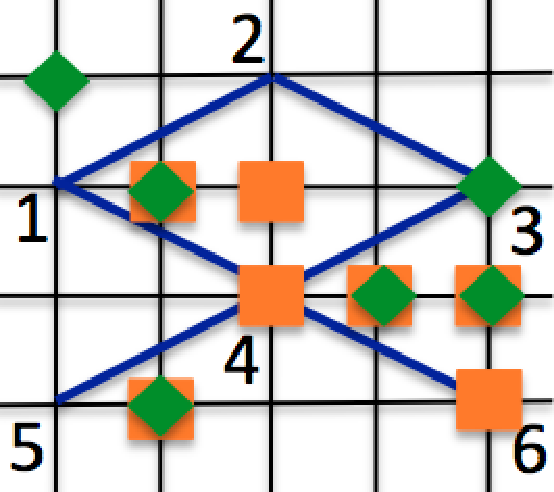}
\caption{
(Color online) Example of bond-occupation. 
Three bonds $\{1\to 4\}$, $\{4\to 3\}$, 
and $\{4\to 6\}$ are occupied in the particle 
configuration, where each circle symbol 
represents a particle in the left-side.
The particle configuration in the left-side
is also generated by using $\beta^+$ and $\beta^-$ as illustrated in the right-side, 
where square and diamond symbols at each site correspond to $\beta^+=1$ and $\beta^-=1$, respectively.}
\label{bond}
\end{figure}

Let $\calB$ be a set of directed bonds written as 
$\{(4k,2l)\to(4k+2,2l\pm1)\}$ or $\{(4k-2,2l+1)\to(4k,2l+1\pm1)\}$, 
where $k,l\in\mathbb{Z}$. 
We say that a bond $\{(n,m)\to(n+2,m\pm 1)\}\in\calB$ 
is occupied when three sites $(n+1,m)$, $(n+2,m)$, 
and $(n+2,m \pm 1)$ are all occupied. 
See the left-side of figure \ref{bond} for an example of occupied bonds. 
Now, if there exists an infinite connected cluster 
(in the sense of a directed percolation) 
of the occupied bonds in $\calB$, we can find frozen
particles in the thermodynamic limit. 
Thus, the problem becomes to be similar to the standard directed percolation in 
a cellular automaton \cite{Hinrichsen}. 
However, the occupation of bond  $\{(n,m)\to(n+2,m+1)\}$ is not independent of
the occupation of bond $\{(n,m)\to(n+2,m-1)\}$ because  
the occupation of the two sites, $(n+1, m)$ and $(n+2, m)$,
influences the two bonds. Due to this effect, it is not straightforward to 
obtain the explicit percolation point.

In order to overcome this difficulty,
we introduce two auxiliary variables, $\beta_i^+$ and $\beta_i^-$,
which take the value $0$ with probability $1-\trho$, and $1$, 
with probability $\trho$. 
Then, one can find that the probability measure of $\sigma$ 
in the initial conditions is the same as that of $(\beta_i^++\beta_i^--\beta_i^+\beta_i^-)_{i\in\Lambda}$ 
if $\trho$ satisfies $\rho=1-(1-\trho)^2$. 
This can simplify the present problem in the following way.
Here, instead of the bond defined by the occupation variable 
$\sigma_i$, we consider the bond defined by the auxiliary 
variables $\beta^{\pm}_i$, as follows. 
As shown in figure \ref{bond}, we say that a bond $\{(n,m)\to(n+2,m+1)\}$ 
is $\beta$-occupied when  $\beta_{n+1,m}^+=\beta_{n+2,m}^+=\beta_{n+2,m+1}^+=1$.
Likewise, $\{(n,m)\to(n+2,m-1)\}$ is called $\beta$-occupied 
when $\beta_{n+1,m}^-=\beta_{n+2,m}^-=\beta_{n+2,m-1}^-=1$.
As $\beta$-occupation for each bond independently occurs 
with the probability $p=\trho^3$, there is a critical value 
\begin{eqnarray}
p_{\rm c}=\trho_{\rm c}^3\simeq 0.644\cdots,
\end{eqnarray} 
above which the directed bond percolation occurs \cite{Hinrichsen}.
It can been seen that if a bond is $\beta$-occupied, 
the bond in the original problem, which is made by the 
relation $\sigma_i=\beta_i^++\beta_i^--\beta_i^+\beta_i^-$, is also occupied.
Thus, an infinite connected cluster of $\beta$-occupied 
bonds indicates an infinite connected cluster of bonds
in the original problem. By using the condition to recover the original problem:
\begin{eqnarray}
\rho_{\rm u}=1-(1-\trho_c)^2,
\end{eqnarray} 
we find that there exists an infinite connected cluster 
(in the sense of a directed percolation) of occupied bonds in 
$\calB$ for  $\rho>\rho_{\rm u}$. 
Thus $\rho_{\rm u} \simeq 0.981\cdots$ is an upper bound.

\section{Conjectures by numerical experiments}
In this section, we numerically try to obtain a greater
lower bound and a smaller upper bound of the transition point $\rho_{\rm c}$
than those given in the previous section.
We also present numerical evidence that the mechanism to characterize 
the {\it spiral} class is not relevant to the freezing transition in the proposed model.
\rd{The key point to obtain useful bounds for $\rho_{\rm c}$ in the model with the filled boundary condition 
is to introduce different boundary conditions from the filled one as explained in the following.}

\subsection{Greater lower bounds}
\begin{figure}
\centering
\includegraphics[width=8cm]{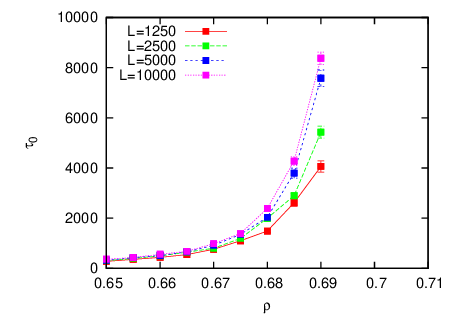}
\caption{(Color online) Relaxation time $\tau_0$ with different system size $L$.}
\label{tau}
\end{figure}

Lower bounds of the transition point are numerically obtained 
by checking whether initial configurations with density $\rho$ 
relax to the equilibrium configuration in the thermodynamic limit. 
\rd{In order to avoid strong finite size effects caused directly from the filled boundary, 
we use the periodic boundary condition on the assumption that 
the transition points obtained in the thermodynamic limit for both of two boundary conditions are identical.
Since we do not have any proof to support this assumption, though it would make sense,
the analysis below should be regarded as a reference for further studies.}
Specifically, we measure the relaxation time $\tau_0\equiv \min \tau'$ such that $H(\sigma(\tau'))=0$ 
as a function of $\rho$.
As shown in figure \ref{tau}, $\tau_0$ increases when $\rho$ is increased. 
Taking into account the size-dependence of $\tau_0$, one might estimate safely $\rho_{\rm c} \ge 0.67$. 
It should be noted that a longer simulation time and also smarter algorithms 
would allow us to obtain better estimations of lower bounds for $\rho_{\rm c}$. 
Here, the rather important point is that a reliable finite-size scaling form of $\tau_0$ 
was not found as far as we studied. One possibility for it might be that $\tau_0$ 
obeys a Vogel-Fulcher type singularity as found in the spiral model \cite{Toninelli2}.

\subsection{Smaller upper bounds and the absence of  the {\it spiral} universality class}
Next we attempt to find smaller upper bounds of the transition point.
In order to investigate the behaviors in high densities, 
\rd{we introduce a half-filled boundary condition that 
particles are filled in the left and right region of the outside of the system, 
while no particles exist in the top and bottom region of the outside of the system.}
Then, we try to observe a percolation of frozen particles in the direction 
from left to right under \rd{this half-filled} boundary condition.
In fact, if such a directed percolation 
occurs in \rd{the case of the half-filled boundary condition}, 
the corresponding frozen particles would be observed in the system with the filled boundary conditions. 
As we will see, one can detect a part of frozen particles \rd{with the filled boundary condition} 
in high densities using the \rd{half-filled boundary condition}.

First, we estimate the probability $P_{L}(\rho)$ 
of finding a percolating frozen cluster, in which there is no $i_x'$ such that 
$\sigma_{i_x'-1,i_y}=\sigma_{i_x',i_y}=\sigma_{i_x'+1,i_y}=0$ for arbitrary $i_y$ 
\rd{for the half-filled boundary condition with} $M=L^{1/{z}}$. We may restrict our investigation to
$0 \le z^{-1} \le 1$ because of the $\pi/2$ rotational symmetry in \rd{the system with the filled boundary condition}. 
Then, for an assumed value of $z$, we determine 
whether a density $\rho_{\rm dp}$ exists such that 
$\lim_{L \to \infty} P_{L}(\rho) = \Theta(\rho-\rho_{\rm dp})$ 
where $\Theta$ is the step function defined before. 
If such a density $\rho_{\rm dp}$ exists, it corresponds to a directed percolation 
point for the system modified with $M=L^{1/{z}}$. 
In this manner, we can determine a set of values $(\rho_{\rm dp}, z)$, and $\rho_{\rm dp}$ provides
an upper bound of the transition point. Practically,
in numerical experiments, for a given value of $z$,
we consider a cross point of the two curves 
$P_{L/2}(\rho)$ and $P_{L}(\rho)$, which is denoted by
$\rho^{L}_{\rm cr}(z^{-1})$. We then check whether
$\rho^{L}_{\rm cr}(z^{-1}) \to \rho_{\rm dp}$ in the limit  $L \to \infty$. 

\begin{figure}
\centering
\includegraphics[width=8cm]{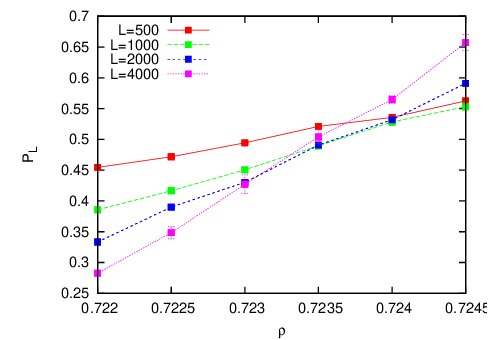}\includegraphics[width=8cm]{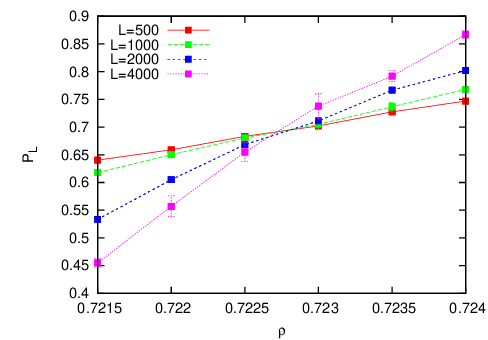}
\caption{(Color online) $P_L(\rho)$ for $z^{-1}=0.63$ \rd{(left)} and $0.68$ \rd{(right)}.}
\label{plm1}
\end{figure}

\begin{figure}
\centering
\includegraphics[width=8cm]{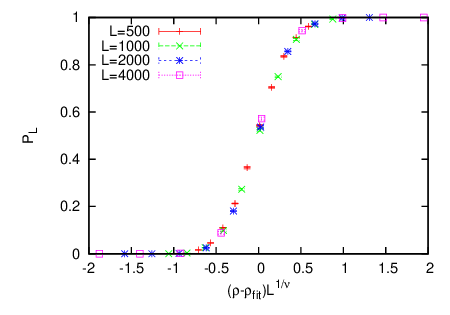}
\caption{(Color online) Finite-size scaling of $P_L(\rho)$ with $z^{-1}=0.63$ and $\nu=1.7338$ 
\cite{Toninelli0}. We used $\rho_{\rm fit}=\rho_{\rm fit}^0\equiv 0.7237$.}
\label{plm2}
\end{figure}

In this way, we consider the possibility of finding a 
convergence point $\rho_{\rm dp}$ with $z^{-1}=0.63$ 
where the criticality of the {\it spiral} class possibly appears. 
As shown in the left side of  figure \ref{plm1}, 
the numerical data up to $L=4000$ suggest that 
$\rho_{\rm cr}^L$ are scattered to judge convergence to a special value 
where the density is changed by an increment of $0.0005$.
At this stage, there are two possibilities; one is 
that there is no convergence point in this procedure 
with $z^{-1}=0.63$, and  the other is that we may 
find the existence of $\rho_{\rm dp}$ by studying 
larger system sizes. The first case means that the 
freezing transition is not connected to the exponent 
$z^{-1}=0.63$. That is, the freezing transition in 
our model does not belong to the {\it spiral} class.
We now elaborate on  the second possibility by employing
a different method, finite-size scaling analysis $P_L(\rho)$.
As shown in figure \ref{plm2}, the finite-size scaling analysis 
with the exponents of the standard directed percolation provides 
reasonable collapsed data with a fitting density 
$\rho_{\rm fit}=\rho_{\rm fit}^0$, which could be a plausible 
estimation for $\rho_{\rm dp}$. 
(Although we tried to collapse the data with different values of 
$\rho_{\rm fit}$, the manner of the collapse with, for example, 
$\rho_{\rm fit}=0.723$ was definitely worse than that for 
$\rho_{\rm fit}=\rho_{\rm fit}^0$.)
However, even in this case, we conjecture that the freezing 
transition in our model does not belong to the {\it spiral} class
by the following reason: As shown in the right side of 
figure \ref{plm1}, we have found much better convergences for 
$z^{-1}=0.68$ with a convergence point $\rho_{\rm dp}$, which is smaller than $\rho=0.723$. 
This means that $\rho_{\rm dp}$ for $z^{-1}=0.68$ may be another smaller upper bound 
than $\rho_{\rm fit}=\rho_{\rm fit}^0$, 
and that the properties at this smaller upper bound
may be characterized by a different value from $z^{-1}=0.63$. 
Thus, this upper bound $\rho_{\rm fit}=\rho_{\rm fit}^0$ 
is not relevant to the transition point of \rd{the model with the filled boundary condition}.
In other words, if we perform similar experiments for the spiral model 
removing particles from the boundaries in a proper way \cite{Toninelli0}, 
where boundaries are $\pi/4$-angle rotated from the present \rd{half-filled} boundaries, 
$\rho_{\rm fit}=\rho_{\rm dp}(z^{-1})$ should hold for $^\forall z$ satisfying $z^{-1}>0.63$ 
because no other percolating (frozen) cluster except for the ones made by 
the standard cellular automaton exist in \rd{the system with such half-filled boundaries}. 
This fact is completely different from the obtained results here.
In sum, for both the two possibilities, these numerical 
results lead to a conclusion that the freezing transition 
in the proposed model does not belong to the {\it spiral} class.

One may point out a possibility that the \rd{diagonally half-filled} boundary
condition that was used for the previous study \cite{Toninelli0}
provides a nontrivial upper bound which is responsible for 
the {\it spiral} class of the transition in the knight model with the filled boundary
condition. Here, the \rd{diagonally half-filled} boundary condition means that
the states of sites along four diagonal lines for the lattice are fixed 
as boundary sites, having particles for one facing pair of two lines and 
no particles for the other pair. However, we have numerically observed
that the percolation density for the present model with the \rd{diagonally half-filled}
boundary condition is very close to $1$ even if it exists,
which is reasonable by considering the properties of
the constraint function. Therefore, the present model with the \rd{diagonally half-filled}
boundary does not have any possibilities to be connected to the
{\it spiral} class of the transition in the model with the filled
boundary condition. Thus, as we have discussed above,
we have considered this possibility for the model with the present \rd{half-filled boundary condition}.

\begin{figure}
\centering
\includegraphics[width=8cm]{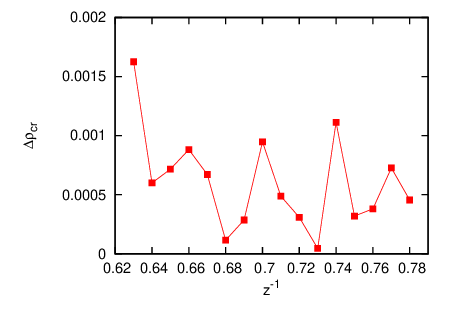}\includegraphics[width=8.0cm]{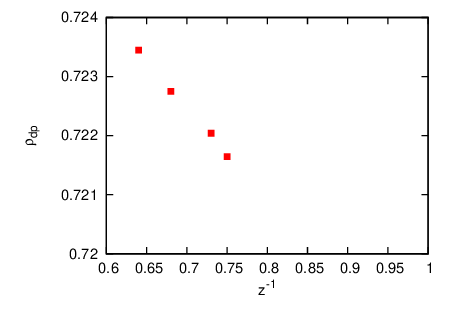}
\caption{(Color online) $\Delta\rho_{\rm cr}$ as a function of $z^{-1}$ \rd{(left)}. 
$\rho_{\rm dp}$ as $\bar\rho_{\rm cr}$ for $z^{-1}=0.64,0.68, 0.73$, and $0.75$ \rd{(right)}.}
\label{plm3}\label{plm4}
\end{figure}

\begin{figure}
\centering
\includegraphics[width=8.0cm]{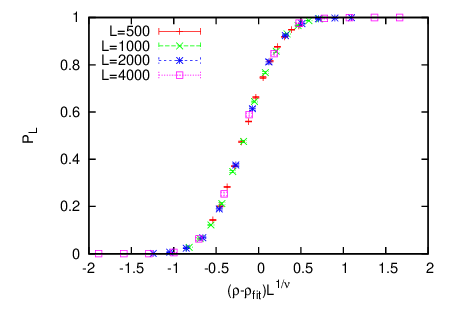}\includegraphics[width=8.0cm]{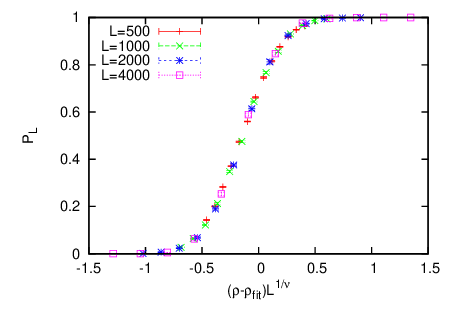}
\caption{(Color online) Finite-size scaling of $P_L$ for $z^{-1}=0.68$. $\nu=1.66$ \rd{(left)} and $\nu=1.7338$ \rd{(right). $\rho_{\rm fit}=0.72275$.}}
\label{F68}
\end{figure}

As a reference, we report tentative numerical results for the dependence of 
$P_L(\rho)$ on $z^{-1}$ from the other viewpoint. We quantitatively investigate the extent of the 
convergence by measuring the average value of the cross points $\bar \rho_{\rm cr} \equiv\sum_{i=1}^{3} 
\rho^{L_i}_{\rm cr}/3$ with $L_{i}=500\times 2^{i}$ 
and the deviation $\Delta \rho_{\rm cr} \equiv  \sqrt{\sum_{i=1}^{3} 
(\rho_{\rm cr}^{L_i}-\bar\rho_{\rm cr})^2/3}$ for each value of $z^{-1}$. 
Further, we estimated $\rho_{\rm cr}^{L_i}$ by approximating $P_L(\rho)$ as a piece-wise linear function. 
When $\Delta \rho_{\rm cr}$ is sufficiently small, 
we expect an obvious convergence of $\rho^{L}_{\rm cr}$ in the limit $L \to \infty$. 
Figure \ref{plm4} presents the numerical results of 
$\Delta \rho_{\rm cr}$ for $0.63 \le z^{-1} \le 0.78$, 
where $\Delta \rho_{\rm cr}$ exhibits an oscillatory behaviour with 
local minima at $z^{-1}=0.64, 0.68, 0.73$, and $0.75$.
Assuming that $\bar\rho_{\rm cr}$  at such local minima 
can be regarded as $\rho_{\rm dp}$, 
we show the $z$-dependence of $\rho_{\rm dp}$ in figure \ref{plm3}. 
Further, this leads us to guess that 
there are such local minima points in $\Delta \rho_{\rm cr}$ at larger values of $z^{-1}$ 
even though it is not easy to obtain them numerically 
due to the longer numerical simulations required.
Therefore, in principle, there exists the smallest upper bound $\rho_{\rm dp*}$ 
that can be obtained by this procedure 
as $\rho_{\rm c}\le \rho_{\rm dp*} \le 0.7217(5) < \rho_{\rm fit}^0$.
The possible interpretation of an oscillatory behaviour in $\Delta \rho_{\rm cr}$ is 
the coexistence of different cellular automaton making distinct directed percolating (frozen) clusters, 
each of which is characterized by different exponents $z$ at different percolation densities $\rho_{\rm dp}$.
This interpretation is not very unreasonable 
because one can easily construct another cellular automaton making frozen clusters at the high densities 
in the model \rd{with the filled boundary condition}, which is different from the cellular automaton introduced in the section 3.2.
\rd{Further, we have performed the finite-size scaling of $P_L$ also for $z^{-1}=0.68$ 
using $\rho_{\rm dp}$ estimated above with $\nu$ as a fitting parameter as shown in the 
left-hand side of figure \ref{F68}. 
At least, the fitting with $\nu=1.66$ supports the existence of a universal curve.
Note that it is not very easy to estimate the best fitting value of $\nu$: 
for example, the fitting with $\nu=1.7338$ would be still reasonable as shown in the 
right-hand side of figure \ref{F68}. 
However, independent of the fitting value of $\nu$,
both of the cases are also consistent with our interpretation
because $z^{-1}=0.68$ is already different from that of the spiral universality.}
\rd{Unfortunately, extensive numerical simulations in order to 
exactly answer whether our interpretation for the numerical results is correct 
remain to be performed in the future.}

\section{Concluding remarks}
The main achievement of this work is the presentation
of a KCM with the \rd{reflection} symmetry that exhibits a freezing transition
on the square lattice. Further, we conjecture by numerical experiments
that the singular behaviour at the freezing transition does 
not belong to the {\it spiral} class
\cite{Toninelli2,Toninelli3}.
Lastly, we comment on future studies in the following text.


First, the characterization of the singular behaviour near the freezing
transition in our model remains to be solved.
More precise estimations for 
the transition point $\rho_{\rm c}$ need to be derived theoretically and numerically.
In particular, it might be interesting if one explicitly constructs
a directed percolation problem related to the behavior near
the transition point. Furthermore, the manner of divergence
of the relaxation time $\tau_0$ should be clarified.
We conjecture that  
$\tau_0$ in our model exhibits a Vogel-Fulcher 
type singularity when $\rho$ approaches $\rho_{\rm c}$ 
from below, in a manner similar to that in the {\it spiral} class.
Presently, we do not have clear evidence for the conjecture, because $\rho_{\rm c}$ 
has not been estimated with sufficient accuracy as yet. After obtaining 
precise estimations for $\rho_{\rm c}$, we will be able to
validate our conjecture by numerical experiments. 
Second, the theoretical analysis of the model on a Bethe lattice
will be performed in order to enhance the understanding of the 
universality of freezing transitions in KCMs. 
A concrete question on Bethe lattices is whether 
the singular behaviour of the freezing transitions 
observed here can be characterized by power-law exponents 
associated with a mode-coupling equation, as discussed in 
the Fredrickson-Andersen model \cite{SBT,Reiter,Kawasaki,Andersen,Sellitto,Ohta,Franz}.
If the answer is yes, since it is different from 
the Vogel-Fulcher type singularity, we will clarify the origin
of the difference between the behaviours of the model on the Bethe 
lattice and on the square lattice.
\rd{Lastly, we would like to mention another dynamics in 
an equilibrium situation with a fixed average of the density \cite{Toninelli3}. 
The freezing transition is unchanged by the definition of the frozen particles 
even in this case. However, strictly speaking, we have not fully understood 
the effects of unfrozen particles in this case, which remain to be clarified in the future.} 
By addressing the points above, we wish to understand how freezing
transitions in KCMs are related, or unrelated, to jamming transitions.

\begin{acknowledgements}
The authors thank H. Tasaki for providing 
us with a basic idea for the proof of the existence of a freezing transition. 
We also  thank C. Toninelli and G. Biroli 
for their discussions on the numerical simulations.
This work was supported by the JSPS Core-to-Core Program 
``International research network for nonequilibrium dynamics of soft matter''.

\end{acknowledgements}



%
%

\newpage
\appendix
\section*{Appendix: Sketch of the proof for the absence of locally frozen particles}
In section 3.1, we assumed the fact that the clusters of frozen 
particles in the bulk are always constrained partly by the boundary sites.
In order to explicitly state such an argument, 
as preliminary, we start with the following definitions
 and will give a sketch of the proof for the argument.

\paragraph{Definition 0 (frozen particles):}
{\it ${\mathcal{F}_{\rm all}}(\sigma)$ denotes a set of all the frozen sites (particles) for 
a given configuration $\sigma$, 
where for $^\forall i\in {\mathcal{F}_{\rm all}}$,  site $i$ is constrained only by 
sites in ${\mathcal{F}_{\rm all}}$ and boundary sites.}
${\mathcal B_{\rm d}}$ denotes a set of boundary sites.

\begin{figure}
\centering
\includegraphics[width=10.0cm]{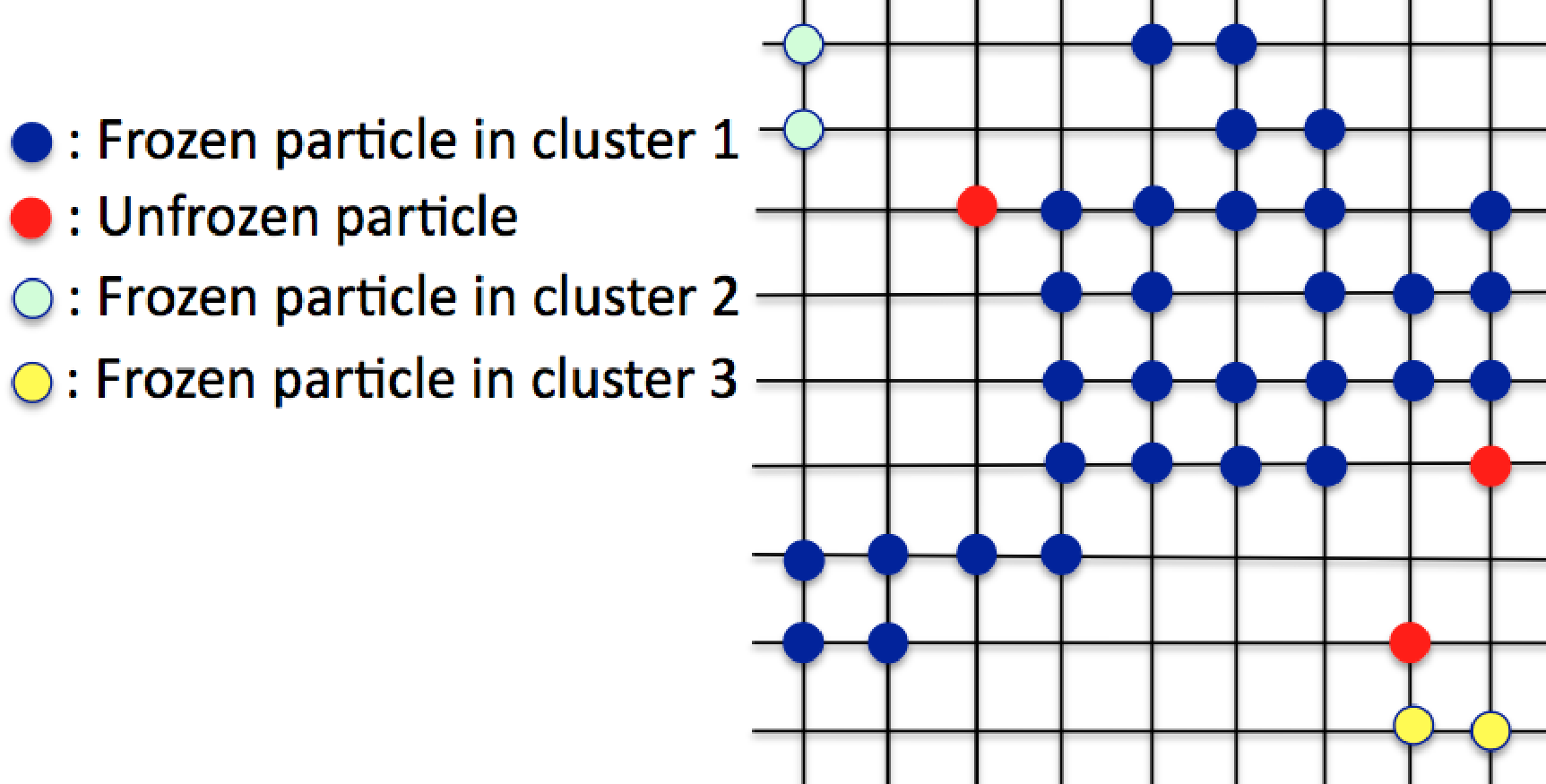}
\caption{(Color online) Frozen particles and clusters. 
The configurations outside of the figures 
are assumed to be consistent.}
\label{frozen}
\end{figure}

\paragraph{Definition 1 (frozen cluster):}
{\it $\mathcal{F}\subset{\mathcal{F}_{\rm all}}$ denotes a set ``frozen cluster'' of frozen particles, 
where for $^\forall i\in \mathcal{F}$, $^\exists j\in{\mathcal F}\cup{\mathcal B}_{\rm d}$ such that $|i-j|\le 3$ 
and (maximum property) for $^\forall k\notin \mathcal{F}$, $k\cup\mathcal{F}$ is not a frozen cluster.} 
See figure \ref{frozen} for helping to imagine the frozen particles and frozen clusters.

\paragraph{Definition 2 (frozen links):}
{\it ${\mathcal L}({\mathcal F})$ denotes a set of ``frozen links'' for ${\mathcal F}$, 
consisting of straight line segments $(i,j)$ between $i$ and $j$ for $^\forall i,j\in \mathcal{F}$.}

\paragraph{Definition 3 (outer sites):}
{\it $\overline{{\mathcal F}}$ denotes a set of outer sites for ${\mathcal F}$, 
where for $^\forall i\in\overline{{\mathcal F}}$, $i$ has a path to a boundary site 
without crossing $^\forall k\in\mathcal{L}({\mathcal F})$.}

\paragraph{Definition 4 (outer links):} 
{\it $\partial\mathcal{L}({\mathcal F})$ denotes a set of outer links, 
where arbitrary points on $k\in \partial\mathcal{L}({\mathcal F})$ has a path 
to a boundary site without crossing $^\forall k\in\mathcal{L}({\mathcal F})$.}

\begin{figure}
\centering
\includegraphics[width=8.0cm]{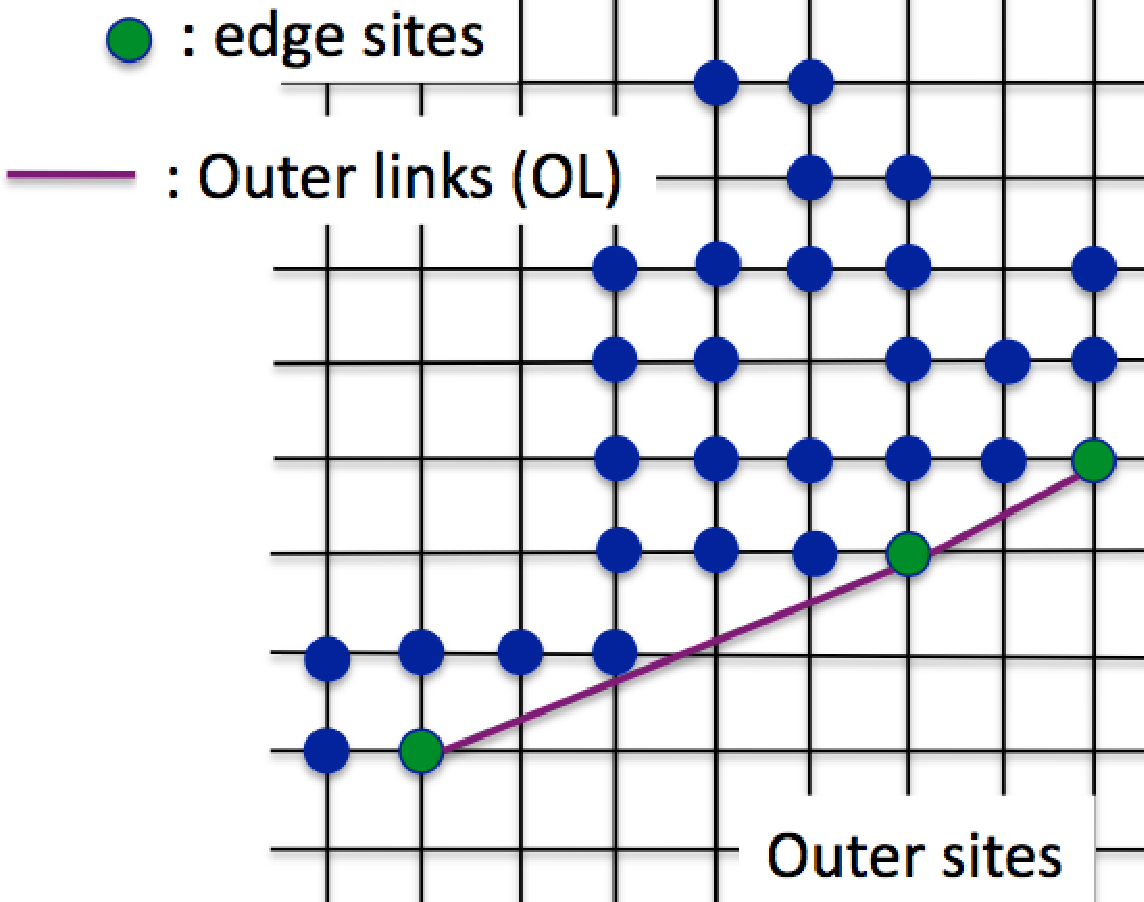}
\caption{(Color online) Outer sites, outer links, and edges sites. 
The configurations outside of the figures 
are assumed to be consistent.}
\label{outer}
\end{figure}

\paragraph{Definition 5 (edge sites):} 
{\it $\partial {\mathcal F}\equiv {\mathcal F}\cap\overline{{\mathcal F}}$ is a set of edge sites.} 
By the definition of edge sites and outer links, for $^\forall (i,j)\in\partial \mathcal{L}(\mathcal{F})$, 
$i,j\in \partial {\mathcal F}$. See figure \ref{outer} for helping to image outer sites, outer links, and edges sites.

\paragraph{}
On the basis of those definitions, one can obtain the following property:
\paragraph{Geometric property 1:}
{\it The figure generated by all the outer links in $\partial\mathcal{L}({\mathcal F})$ 
are locally convex toward $\mathcal{F}$.}

\paragraph{}
Assume that the figure generated by $(i,j),(j,k)\in\partial \mathcal{L}(\mathcal{F})$ 
is not convex toward $\mathcal{F}$. Then, one can construct another outer link connecting $i$ and $k$, 
and $(i,j),(j,k)\notin\partial \mathcal{L}(\mathcal{F})$ by the definition of the outer links.
This contradiction immediately leads to {\it Geometric property 1}.

\paragraph{}
By the local convexity in {\it Geometric property 1}, one may immediately conclude the following property:
\paragraph{Geometric property 2:}
{\it The figure generated by all the outer links in $\partial\mathcal{L}({\mathcal F})$ 
is a convex polygon or a convex polyline toward $\mathcal{F}$.  } 

\paragraph{}
Finally, we explicitly state about the absence of locally frozen particles in the following.
\paragraph{Statement:} 
{\it There exist frozen particles constrained by the boundary sites for arbitrary frozen clusters: 
For $^\forall {\mathcal F}\neq\emptyset$, $^\exists i\in {\mathcal F}$ such that $\min_{j\in{\mathcal B}_{\rm d}}|i-j|\le 3$.}

\begin{figure}
\centering
\includegraphics[width=4.5cm]{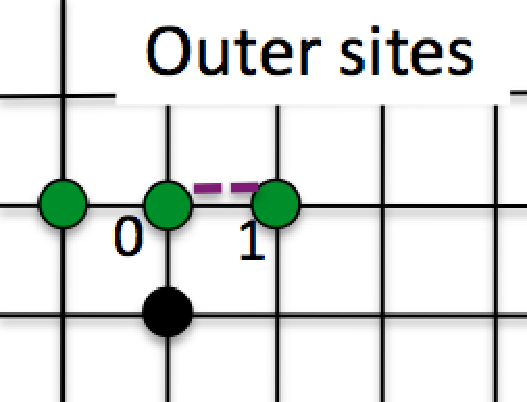}
\caption{(Color online) The case with $3$ particles at sites in ${\rm B}_0$. 
In this case, the slope of outer links $(0,1)\in\partial\mathcal{L}({\mathcal F})$ 
clearly does not change, comparing to the previous outer link. As far as the edge sites 
continue to have three particles at the nearest neighbors, this situation does not change.}
\label{pattern1}
\end{figure}

\paragraph{Sketch of the proof:} 
Assume that there exists a frozen cluster ${\mathcal F}$ 
such that $^\forall i\in {\mathcal F}$, $\min_{j\in{\mathcal B}_{\rm d}}|i-j|> 3$. 
Then let us consider what kinds of configuration could appear near edge sites 
in $\partial {\mathcal F}$. Remembering the {\it maximum property} of frozen clusters, 
one can easily find that each site in $\partial {\mathcal F}$ 
has at least two particles, but less than four particles at the nearest neighbors. 
The case with three particles is illustrated in figure \ref{pattern1}, and 
it can been seen that one immediately has to consider the case with two particles.

\begin{figure}
\centering
\includegraphics[width=5.5cm]{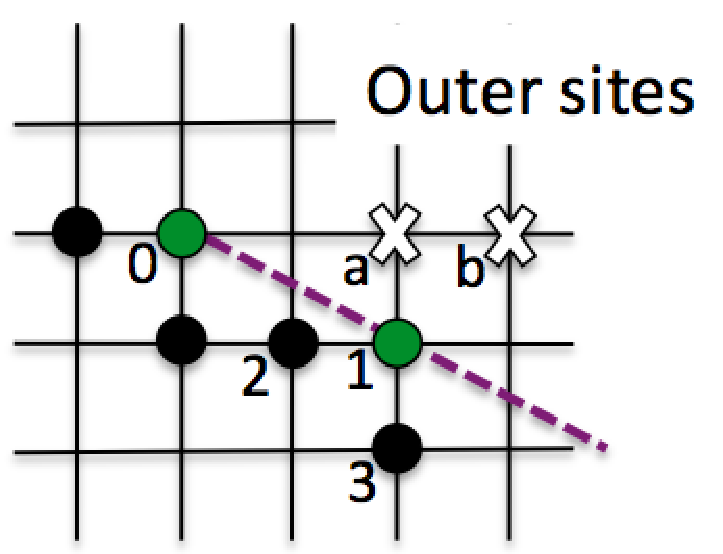}
{\ }{\ }{\ }{\ }{\ }{\ }{\ }{\ }{\ }{\ }{\ }{\ }
\includegraphics[width=5.5cm]{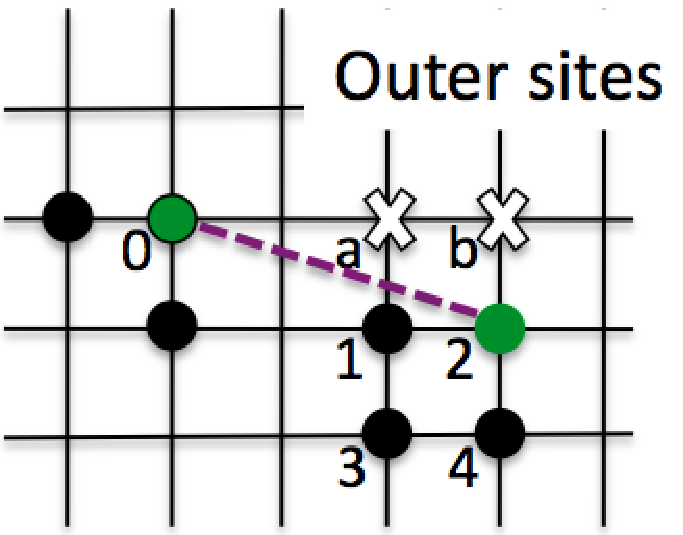}
\caption{(Color online) The situations where an edge site $0$ has two particles in ${\rm B}_0$ 
and a particle at site $1$ makes site $0$ constrained ($a$ and $b$ are outer sites in $\overline{{\mathcal F}}$). 
It is necessary to consider these rather particular situations 
because even if we consider the cases where site $0$ is constrained by the other sites 
such as sites $a$ and $b$, one clearly has to consider these situations again in order to make a polygon .
(left) Assume that site $1$ is an edge site in $\partial{\mathcal F}$. 
In this case, the situation that there are particles at sites $2$ and $3$ 
is only the possibility to be consistent with the definitions made in the text.
Finally, the edge site $1$ has the same situation as that of site $0$. 
Therefore, The slope of outer link $(0,1)$ cannot become sharper 
than that of the next outer link $(1,k)\in\partial\mathcal{L}({\mathcal F})$. 
(right) Assume that site $1$ is not an edge site. 
In this case, there must exist a particle at site $2$. 
Assume that site $2$ is an edge site in $\partial{\mathcal F}$. In this case, 
the situation that there are particles at sites $1$ and $4$, is only the possibility to be consistent 
with the definitions made in the text. Finally, the edge site $2$ has the same situation as that of site $0$, 
and the slope of outer link $(0,2)$ is not sharper than that of outer link $(0,1)$. 
Finally, if one consider the last possibility that site $2$ is not an edge site, 
the slope of the outer link $(1,k)\in\partial\mathcal{L}({\mathcal F})$ cannot be sharper 
than that of outer link $(0,2)$ by the definition of the outer links. 
Concretely, this is because if the slope of outer link $(1,k)$ would be sharper than that of outer link $(0,2)$, 
site $2$ would be an edge site, which leads to the contradiction.}
\label{pattern2}
\end{figure}

Thus, we focus on the case with two particles at the nearest neighbor of an edge site. 
According to {\it Geometric property 1}, one can illustrate two possible and nontrivial configurations as 
illustrated in figure \ref{pattern2}. Concretely, we first pick up an edge site $0$ in $\partial {\mathcal F}$, 
and consider the slope of outer links $(0,k)\in\partial \mathcal{L}({\mathcal F})$ to curve downward 
in order to make a polygon, without loss of generality because one has to consider these cases in the end at the latest. 
However, it turns out that the slope of outer links $(0,k)\in\partial \mathcal{L}({\mathcal F})$ 
in both configurations cannot be changed to make any polygons (see the caption in figure \ref{pattern2}).
Note that the cases where there is a particle at site $a$ or $b$ to make site $0$ constrained are also possible, 
but the slope of the outer link $(0,k)\in\partial \mathcal{L}({\mathcal F})$ 
does not change downward at all. As explained in the caption of figure \ref{pattern2},
 those cases are enough to consider impossiblity for the outer links to make them 
being a polygon base on the assumption we have made.
Therefore, one may conclude that the figure generated by 
all the outer links in $\partial\mathcal{L}({\mathcal F})$ is not a polygon on the assumption we have made.

Therefore, another possibility is that the figure generated by 
all the outer links in $\partial \mathcal{L}({\mathcal F})$ is a convex polyline 
according to {\it Geometric property 2}.
However, in this case,  the tip of the line has to be constrained by a boundary site 
because the tip of the line should have at least one more particle at the nearest neighbors by the definition. 
Thus, one may conclude there are contradictions, leading to {\it Statement}. 

\end{document}